\documentclass[aps,prl,twocolumn,preprintnumbers]{revtex4}

\usepackage[dvips]{color}
\usepackage[normalem]{ulem}
\usepackage{amsmath}
\usepackage{enumerate}
\usepackage{amsfonts}
\usepackage{epsfig}

\newcommand{\be}{\begin{equation}}
\newcommand{\ee}{\end{equation}}
\newcommand{\bea}{\begin{eqnarray}}
\newcommand{\eea}{\end{eqnarray}}

 \newcommand{\bln}{\begin{align}}
\newcommand{\eln}{\end{align}}
\newcommand{\bst}{\begin{split}}
\newcommand{\est}{\end{split}}
\newcommand{\bi}{\begin{itemize}}
\newcommand{\ei}{\end{itemize}}
\newcommand{\ben}{\begin{enumerate}}
\newcommand{\een}{\end{enumerate}}

\def\ov{\over}
\def\le{\left}
\def\ri{\right}

\def\lam{{\lambda}}
\def\Lam{{\Lambda}}
\def\al{{\alpha}}

\def\det{{\rm det}}

\def\NN{{\cal N}}
\def\th{{\theta}}

\def\Om{{\Omega}}
\def \th{{\theta}}

\def \lam {\lambda}
\def \om {\omega}

\def\sig{{\sigma}}

\def\apr{{\alpha'}}
\newcommand{\p}{\partial}

\newcommand\Ga{{\ensuremath{{\Gamma}}}}

\def\lam{{\lambda}}

\def\eeq{\end{equation}}

\newcommand\sN{{\ensuremath{{\mathcal N}}}}

\begin{document}

\title {Meson widths from string worldsheet instantons}

\preprint{MIT-CTP 3959}

\author{Thomas Faulkner and Hong Liu}

\affiliation{Center for Theoretical Physics,
Massachusetts
Institute of Technology,
Cambridge, MA 02139
}

\begin{abstract}

We show that open strings living on a D-brane which lies
outside an AdS black hole
can tunnel into the black hole through worldsheet instantons.
These instantons have a simple interpretation in terms of
thermal quarks in the dual Yang-Mills (YM) theory.
As an application we calculate the width of a meson in a strongly coupled quark-gluon
plasma which is described holographically as a massless mode on a
D7 brane in $AdS_5 \times S_5$.
While the width of the meson is zero to all orders in the $1/\sqrt{\lam}$ expansion with $\lam$ the 't Hooft coupling, it receives non-perturbative contributions in $1/\sqrt{\lam}$ from worldsheet instantons. We find that the width increases quadratically with momentum at large momentum and comment on potential phenomenological implications of this enhancement for heavy ion collisions.
We also comment on how this non-perturbative effect has
important consequences for
the phase structure of the YM theory obtained in the classical gravity limit.

\end{abstract}

\maketitle
\newpage

A heavy quarkonium bound state, like $J/\psi$ or $\Upsilon$,
which finds itself in the quark-gluon plasma (QGP),
becomes increasingly unstable and eventually dissociates at sufficiently high
temperatures. On the one hand this can be attributed to the weakening attraction
between a heavy quark and anti-quark in the bound state due to color screening of the medium~\cite{Matsui:1986dk}. On the other hand the bound state can be broken up from
collisions with the deconfined quarks and gluons in the medium~\cite{Kharzeev:1994pz}.
Given that the QGP at RHIC and LHC is likely strongly coupled,
understanding such medium effects on the propagation and dissociation
of heavy quarkonia
presents nontrivial theoretical challenges which must be confronted
in order to interpret experimental data on quarkonium suppression.

Interesting insights have recently been made into this problem from
strongly coupled model gauge theories using the AdS/CFT correspondence.
AdS/CFT-based methods are powerful at attacking questions of
dynamical origin, such as how the motion of quarkonia relative to the medium
affects their various properties.
The simplest example of the correspondence is provided by the duality between
${\cal N}=4$ $SU(N_c)$ super Yang-Mills (SYM) theory and string theory in
$AdS_5\times S_5$~\cite{AdS/CFT}. Based on a calculation of the potential between a heavy external quark and antiquark pair moving in the strongly coupled hot $\sN = 4$ plasma, it has been argued in~\cite{Liu:2006nn} (see also~\cite{Peeters:2006iu}) that the dissociation temperature $T_{\rm diss}$ of a heavy quarkonium  {\it decreases} with their velocity $v$ relative to the medium as $T_{\rm diss} (v) \approx (1-v^2)^{1 \ov 4} T_{\rm diss} (v=0)$. Such a velocity scaling, which can be heuristically understood as due to increased screening from the boosted medium, could provide a significant additional source of quarkonium suppression at nonzero transverse momentum in heavy ion collisions~\cite{Liu:2006nn}.

Rather than drawing inferences from the heavy quark potential, it is also possible to directly study the propagation of mesons in a strongly coupled plasma.
While  $\NN=4$ SYM theory itself does not contain dynamical mesons, one can obtain a closely related theory which does contain mesons by adding to it $N_f \ll N_c$ fundamental ``quarks'', which corresponds to adding some D7-branes to $AdS_5\times S_5$ in the gravity picture~\cite{Karch:2002sh}.
It was found in~\cite{Mateos:2007vn,Ejaz:2007hg} that
 meson dispersion relations are dramatically modified by the plasma and in particular, there exists a limiting velocity $v_c (T)< 1$, which decreases with increasing temperature. The existence of a subluminal limiting velocity is consistent with the velocity-enhanced screening obtained from the heavy quark potential, as when $v > v_c (T)$ the quark and anti-quark are completely screened and no bound states are possible.

For a more complete understanding of the dissociation of mesons
one needs to study their widths.
We will be particularly
interested in the momentum dependence of the widths.
This has not been possible within the classical gravity approximation
developed so far. In this approximation, the mesons are stable (i.e. they have zero width) for $T$ smaller than a dissociation temperature $T_{\rm diss}$, but completely disappear for $T > T_{\rm diss}$~\cite{Kruczenski:2003be,Hoyos:2006gb,Mateos:2007vn}.
The approximation also has another important drawback:
the densities of quarks and antiquarks are identically zero for a range of temperatures and chemical potentials~\cite{Mateos:2007vc} even though they should obey the standard thermal distribution.

In this paper, we discuss a novel tunneling effect on the string worldsheet
which gives rise to nonzero quark densities and meson widths for $T < T_{\rm diss}$.
This enables us to calculate explicitly the momentum
dependence of the width of a meson in a strongly coupled QGP. We find that the width  increases quadratically with momentum at large momentum.

At finite temperature, $\NN=4$ SYM theory can be described by a string theory in
the spacetime of a black hole in AdS$_5 \times S_5$, whose metric can be written as
 \be \label{adsM}
 {ds^2} = - fdt^2 + {1 \ov f} dr^2 + {r^2 \ov R^2} d \vec x^2 + R^2
 d \Om_5^2
 \ee
where $f = {r^2 \ov R^2} \le(1-{r_0^4 \ov r^4} \ri)$, $\vec x = (x_1, x_2, x_3).$
$d \Om_5^2$ is the metric on a unit five-sphere $S_5$ which can be written as
  \be \label{s5m}
d \Om_5^2 = d \th^2 +  \sin^2 \th d\Omega_3^2 + \cos^2 \th d \phi^2, \quad \th \in \le[0, {\pi \ov 2} \ri]\
 \ee
with $d \Om_3^2$ the metric for a three-sphere $S_3$.
The string coupling $g_s$ is related to the Yang-Mills coupling $g_{YM}$ by $g_{s} = 4 \pi g_{YM}^2$ and the curvature radius $R$ is
related to the 't Hooft coupling $\lam = g_{YM}^2 N_c$ by ${R^2 \ov \apr} = \sqrt{\lam}$.
The perturbative $g_s$
and $\apr$ expansions in the bulk string theory are related to the $1/N_c$ and
 ${1 \ov \sqrt{\lam}}$  expansions in the Yang-Mills theory respectively.
The temperature $T$ of the YM theory
is given by the Hawking temperature of the black hole,  $T = \frac{r_0}{ \pi R^2}$.
Adding $N_f$
fundamental ``quarks''
can be described in the dual string theory
by adding
$N_f$ D7-branes in (\ref{adsM})~\cite{Karch:2002sh}.
A fundamental ``quark'' in the YM theory can be described by an open string with one end on the D7-branes and the other end on the black hole. Open strings with both ends on the D7-branes can be considered as ``bound states'' of a quark and antiquark, thus describing meson-type excitations in the YM theory.

\vspace{-0.2cm}

\begin{figure}[h!]
\centerline{\hbox{\psfig{figure=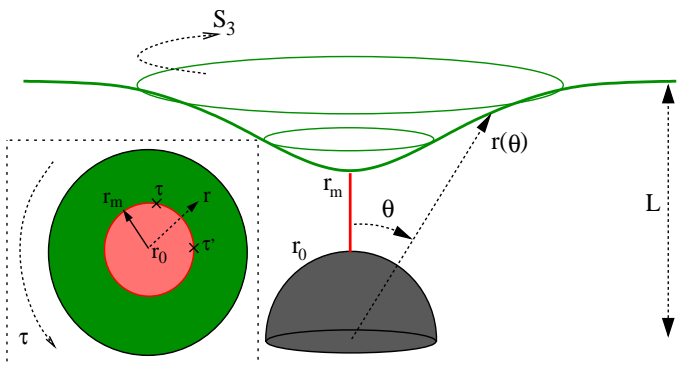,width=9cm,
height=5.6cm}}}
\vspace{-1cm}
\caption{An embedding of the D7 brane (green) in the $AdS_5 \times S_5$ black hole
geometry for $T<T_{\rm{diss}}$ which lies entirely outside the black hole.
Inset: the Euclidean $r-\tau$ plane at $\th=0$ showing a world-sheet instanton (red)
connecting the tip of D7 brane $r=r_m$ to the horizon at the center of the disk
$r=r_0$.
\label{fig:1} }
\end{figure}

\vspace{-0.2cm}

We now briefly outline the standard procedure for obtaining the meson spectrum~\cite{Kruczenski:2003be}.
We will take $N_f=1$, $N_c \to \infty$, and $\lam$ large but finite throughout the paper.
The D7-brane can be chosen to lie along the directions $\xi^\al = (t, \vec x , \Om_3, \th)$
and using the symmetries of the problem the embedding in the two remaining transverse
directions can be taken as $\phi(\xi^\al) = 0$ and $r(\xi^\al) = r(\th)$.
At the lowest order in the $\apr$ expansion, $r(\th)$ can be determined by extremizing the Dirac-Born-Infeld (DBI) action of the D7-brane with the boundary condition $r (\th) \cos \th|_{\th \to {\pi \ov 2}} \to L$, where $L$ is related to the mass $m_q$ of a quark in the Yang-Mills theory as $m_q = {L \ov 2 \pi \apr}$.
For $T$ smaller than some $T_{\rm diss}$,
$r (\th)$ has the form shown in Fig.~\ref{fig:1}. The brane is closest to the black hole at $\th=0$, where there lies a 4-dimensional subspace spanned by $(t, \vec x)$ since here the $S_3$ in (\ref{s5m}) shrinks to a point.
 Denoting $r_m \equiv r(\th=0) > r_0$, the shortest open string connecting the D7-brane to the horizon has a mass in the YM theory
 \be \label{quM}
 m_q^{(T)} = {r_m - r_0 \ov 2 \pi \apr} = \sqrt{\lam} T {\Lam_m -1 \ov 2}, \quad
 \Lam_m = {r_m \ov r_0} \
 \ee
Note that $\Lam_m$ is a dimensionless number of $O(\lam^0)$ determined by the ratio $L/r_0$,
and $m_q^{(T)}$ can be interpreted as the effective mass of a quark at temperature $T$.

The mesons corresponding to massless fluctuations on the D7-brane can be found by
solving the linearized equations resulting from expanding the DBI action around the embedding. For example, the quadratic action for the fluctuation $\chi^\phi (\xi^\al)$ of the location of D7-brane in the $\phi$ direction  can be written as
 \be \label{chia}
 S_{DBI} [\chi^\phi] = - {\mu_7  \ov 2} \int d^8 \xi \, \sqrt{-g}  \, G_{\phi \phi} \,
 g^{\al \beta} \p_\al \chi^\phi \p_\beta \chi^\phi
 \ee
where $\mu_7 = {1 \ov (2 \pi)^7 g_s \al^{'4}}$ is the tension of the D7-brane, $G_{\phi \phi} = R^2 \cos^2 \th$,
and $g_{\al \beta}$ denotes the induced metric on the D7-brane.
 Writing $\chi^\phi = e^{-i \om t + i \vec k \cdot \vec x} Y_l (\Om_3) \psi (\th)$, the equation of motion for $\psi$
 can be written as
 \be \label{SchE}
 \hat{H} (\vec k, l) \psi (\th)= \om^2 \psi (\th)
 \ee
 where $\hat{H}(\vec k, l)$ is a second order differential operator in $\th$
and $Y_l$ are spherical harmonics on the $S_3$.
 For  a given $\vec k, l$, $\hat{H} (\vec k, l)$ has only discrete eigenvalues $\om_n^2$ labeled by an integer $n$, giving rise to dispersion relations $\om = \om_n (\vec k, l)$, all of which have zero width.
In particular, the meson masses are of order $M = {2\sqrt{2} L \ov R^2} = {4 \sqrt{2} \pi m_q \ov \sqrt{\lam}}$. Since $M$ is parametrically smaller than $m_q$ in $\sqrt{\lambda}$, the mesons have a large binding energy, given by $2m_q^{(T)}$.
There exists a temperature $T_{\rm{diss}} = 0.122 M$, beyond which
the D7 brane  falls into the black hole
and mesons cease to exist as well-defined quasi-particles~\cite{Hoyos:2006gb,Mateos:2007vn}.

We stress that the zero-width conclusion only depends on the topology of the embedding in Fig.~\ref{fig:1}.
Since mesons can only dissociate by falling into the black hole,
when the D7-brane lies above the black hole horizon the mesons are necessarily stable.
Given that the brane embedding and the background geometry are smooth, including higher order perturbative corrections in $\apr$ should not change the topology of the brane embedding if the distance between the brane and the horizon is parametrically larger than the string scale.
This implies that the widths of mesons should remain zero to all orders in the perturbative ${1 \ov \sqrt{\lam}}$ expansion.

One can also turn on a quark chemical potential $\mu < m_q^{(T)}$ in the boundary theory by setting $A_t = \mu $, where $A_t$ is the time component of the gauge field on the D7-brane~\cite{Mateos:2007vc,Karch:2007br}. Since the DBI action and its  higher order $\apr$ corrections contain only derivatives of $A_t$, the D7-embedding and the meson spectrum are not modified by turning on the constant mode of $A_t$. Thus, the meson widths and the net quark density remain zero to all orders in the $\apr$ expansion even at a finite chemical potential~\footnote{Above $\mu > m_q^{(T)}$ a new phase, where the D7 brane
falls into the horizon, is thermodynamically preferred. This phase
exhibits both meson widths and finite quark density
\cite{Erdmenger:2007ja,Myers:2008cj,Karch:2007br,Mateos:2007vc}.
}.

The above conclusions can be further illuminated by simple thermodynamic
reasoning.
From (\ref{quM}), $\beta m_q^{(T)} \propto \sqrt{\lam}$, the quark (or anti-quark) density, being proportional to $e^{-\beta m_q^{(T)} \pm \beta \mu}$, is then exponentially suppressed in $\sqrt{\lam}$ for $\mu < m_q^{(T)}$. Similarly, since the binding energy $E_{BE}$ of a meson is $2 m_q^{(T)}$, thermal effects which destabilize the mesons
are also exponentially suppressed in $\sqrt{\lam}$. Thus the meson widths and quark densities are not visible in the perturbative expansion in $1/\sqrt{\lam}$ and can only have non-perturbative origins on the worldsheet.

There are indeed non-perturbative corrections in $\apr$ which effectively change the topology of the D7-brane embedding and generate non-vanishing meson widths and quark densities. To see this it is more convenient to analytically continue (\ref{adsM}) to Euclidean signature with $t \to - i \tau$. Then the $r-\tau$ plane has the topology of a disk. The angular direction
$\tau$ has a period given by the inverse temperature $\beta$.
The center of the disk is located at $r=r_0$. Open strings on the D7-brane are described by worldsheets with the topology of a disk whose boundary lies on the D7-brane.
Denoting the worldsheet coordinates as $\rho \in [0,1]$ and $\sig \in [0, 2 \pi)$,
the worldsheets separate into different topological sectors corresponding
to the winding number $m$ of the target space disk $(r, \tau)$ wrapping around the worldsheet disk $(\rho,\sig)$.  The DBI action arises
from the sector of trivial winding number $m=0$, in which $(\rho, \sig)$ is mapped to a single point on the D7-brane. In all the other (nontrivial) sectors, the string worldsheet is mapped to the region in the $r-\tau$ plane from the location of the D7-brane all the way to the horizon $r=r_0$ (see inset of Fig.~\ref{fig:1}). When analytically continued back to the Lorentzian
signature, such a worldsheet describes a tunneling process in which a tiny neck is generated between the D7-brane and the black hole horizon.
As a result mesons can leak through the tiny neck into the black hole and dissociate.

As an illustration, we now calculate the contributions from $m=\pm 1$ sectors to the quark density and the widths of mesons in (\ref{chia}).
We will only be interested in the lowest order term in the $\apr$ expansion.
The relevant spacetime effective action for the D7-brane
 can be obtained from the worldsheet path integral~\cite{Fradkin:1985qd}
 \be \label{EuP}
  S_{E} [\chi^\phi] = \int_{\rm disk} DX \, e^{- I [X] + \oint_{\rho=1} d \sig \, \mu {d \tau \ov d \sig} - I_B [\chi^\phi]}
 \ee
where  $X= (\xi_\al, r, \phi)$ denotes collectively the worldsheet fields. $I[X]$ is the worldsheet action, which for our purpose can be taken
 to be the Nambu-Goto action for a string propagating in~(\ref{adsM})
\begin{equation}
I[X] ={1 \over 2 \pi \alpha'} \int d \sig d\rho  \, \sqrt{ \det h_{a
b}}
\label{NGaction}
 \end{equation}
with $h_{ab} = G_{M N} \partial_a X^M \partial_b X^N$
 the induced metric on the worldsheet and $G_{MN}$ the Euclidean version
 of the metric (\ref{adsM}).
The second term in the exponential of (\ref{EuP}) corresponds to turning on  $A_\tau = - i \mu$ which gives in the Euclidean theory a nonzero (quark) chemical potential $\mu$ in the boundary YM theory.
$I_B  [ \chi^\phi] = \oint_{\rho=1} d \sig \,
{ R^2  \ov 2 \pi \alpha'} \chi^\phi \p_\rho \phi$
is the boundary action which couples the worldsheet to $\chi^\phi (\xi^\al)$.
We have suppressed any dependence on spacetime and world sheet fermions.
The integral (\ref{EuP}) can be evaluated using the saddle point approximation in each topological sector~\cite{Dine:1986zy}, i.e.
 $S_{E} = S_{m=0} + S_{m=+1} + S_{m=-1} + \cdots$ with $S_{m=0} = S_{DBI}$.

For $m = \pm 1$, (\ref{NGaction}) has a classical solution given by
 \be \label{n=1S}
 \tau = \pm {\beta \ov 2 \pi} \sig, \quad r = \hat r (\rho), \quad \th = 0, \quad \phi=0, \quad \vec x= \vec x_0
 \ee
where $\vec x_0$ is an arbitrary constant position vector and $\hat r(\rho)$ is chosen so that
$\hat r (1) = r_m$ and $f d \tau^2 + {1 \ov f} d\hat r^2 \propto d \rho^2 + \rho^2 d \sig^2$.
Eq. (\ref{n=1S}) represents the worldsheet of a string connecting the tip of the brane to the horizon with the $\pm$ sign corresponding to opposite orientations. It has a classical action $I_{\pm} = \beta m_q^{(T)}$ where $m_q^{(T)}$ is the effective quark mass introduced in (\ref{quM}). One can readily verify that (\ref{n=1S}) minimizes the action and satisfies the right boundary conditions at the D7-brane. Note that there are only three bosonic zero modes in (\ref{n=1S})~\footnote{In contrast, the $n=0$ sector has eight zero modes corresponding to all directions on the D7-brane. There are also no fermionic zero modes here.}, since it costs energy to move away from $\th =0$ and the worldsheet time $\sig$ now coincides with $\tau$. With $\chi^\phi$ set to zero, Eq. (\ref{EuP}) then yields
 \be \label{qudD}
 S_{m = \pm 1} =
e^{- \beta m_q^{(T)}} e^{\pm \mu \beta}
{1 \ov g_s} D  V_3  
 \ee
where the
$e^{\pm \mu \beta}$ arises from the second term in the exponential of (\ref{EuP}), $V_3$ is the spatial volume from integrating over the three zero modes in (\ref{n=1S}), and the ${1 \ov g_s}$ factor arises because we are evaluating
the disk path integral. $D$ is a real number coming from Gaussian integration around the saddle point (\ref{n=1S}) (including worldsheet fermions) whose sign we will fix from physical requirements.
Identifying the Euclidean action with $\beta F (\beta, \mu)$ where
$F (\beta, \mu)$ is the free energy, equation (\ref{qudD}) leads to a net
quark {\it charge} density $- {2 D \ov g_s} e^{- \beta m_q^{(T)}}
\sinh \beta \mu$,
which in turn requires that $D$ should be
{\it negative}~\footnote{There are a few other indications that $D$
should be negative. The instanton action also induces a tadpole for
the $r$ component of the transverse fluctuations. One finds for $D$
negative the tadpole pulls the brane toward the horizon as required
by physical consistency. Also only for $D$ negative, do the meson widths
we calculated below have the correct sign.}.
It is natural to interpret $S_{m=\pm 1}$ as the contributions from quarks
and anti-quarks separately: $S_{m=\pm 1} = -  n_{\pm} V_3$, which from (\ref{qudD}) 
leads to a quark and antiquark number density given by
$n_{\pm} = e^{- \beta m_q^{(T)}} e^{\pm \mu \beta} {1 \ov g_s} (-D)$.
Note that $n_{\pm} \propto 1/g_s \propto N_c$ since quarks come in
an $N_c$-multiplet.

In our derivation of (\ref{qudD}) we have assumed the embedding of the D7-brane is given by that determined by the DBI action. This is justified for $\mu < m_q^{(T)}$ since the correction to the DBI action is exponentially small. 
When $\mu \geq m_q^{(T)}$, the backreactions from instantons become large
and the embedding of Fig.~\ref{fig:1} cannot be trusted.

The nonzero quark densities for nonzero $\mu < m_q^{(T)}$  have important implications for the phase structure of the theory. As discussed in~\cite{Mateos:2007vc} (see also~\cite{Karch:2007br}) based on the analysis of the DBI action (which corresponds to $\lam = \infty$), at low temperature there is a phase transition in chemical potential at which the net quark charge density becomes nonzero. Our results strongly indicate this phase transition (at nonzero temperature) is smoothed to a crossover at any finite value of~$\lam$.

To find the widths of the mesons described by~(\ref{chia}), we need to
compute  (\ref{EuP}) to quadratic order in $\chi^\phi$. For simplicity, we will restrict to the $l=0$ mode on the $S_3$.
Near $\th =0$, the worldsheet action for $\phi$ is given by ${R^2 \ov 4 \pi \apr} \int d\rho d \sig \, (\p \phi)^2$, which is free. The path integral is then straightforward to compute and yields for $S_{m=\pm 1} $
  \begin{equation} \label{noac}
-{R^2 n_{\pm} \ov 2 \pi \apr \beta^2} \int \! d^3 x_0  {d \tau d \tau' } \,\le(  \chi^\phi (\tau, \vec x_0)
  \tilde G_D (\sig, \sig')  \chi^\phi (\tau', \vec x_0) \ri)_{\th=0}
\end{equation}
where $\tilde G_D (\sig,\sig') =  \lim_{\rho \to 1, \rho' \to 1} \p_{\rho} \p_{\rho'} G_D (\rho, \sig; \rho', \sig')$ with $G_D (\rho, \sig; \rho', \sig')$ the Dirichlet propagator for a canonically normalized massless field on the unit disk and $\sig = {2 \pi \ov \beta} \tau$.
  Note that (\ref{noac}) only depends on the
 value of $\chi^\phi$ at the tip of the brane and is nonlocal in the Euclidean time direction.

Treating (\ref{noac}) as a small perturbation to (\ref{chia}), one can compute the corrections to the Euclidean two-point function of the (meson) operator dual to $\chi^\phi$ in the boundary YM theory, from which the corrections to the Lorentzian retarded function can be found by analytic continuation. One finds that the poles of the retarded function now obtain a nonzero imaginary part.
Alternatively one can directly obtain the Lorentzian counterpart of (\ref{noac}) by analytically continuing
the worldsheet disk to Lorentzian signature with $\sig = i \eta
= i 2 \pi t/\beta $, which gives the part of Rindler spacetime $ds^2 = d \rho^2 - \rho^2 d \eta^2$ with $\rho \leq 1$.
The Lorentzian spacetime effective action can be obtained using the Schwinger-Keldysh contour,
giving the Lorentzian equation of motion~\footnote{Eq.~(\ref{eomC}) can also be obtained directly from the Euclidean action (\ref{noac})
using the following general prescription: write down the equation of motion following from the Euclidean action; replace the Euclidean worldsheet time by the Lorentzian worldsheet time and the Euclidean propagator by the corresponding retarded propagator in the Lorentzian signature.}
\be \label{eomC}
\partial_\alpha \left(\sqrt{-g} G_{\phi \phi} \partial^\alpha \chi^\phi  \right)
= - {R^2 n_\pm \delta(\theta) \ov \mu_7  \pi \apr \beta^2} \int dt'\,  \tilde{G}_R(\eta-\eta') \chi^\phi(t')
\ee
where $\tilde G_R (\eta-\eta') =  \lim_{\rho \to 1, \rho' \to 1} \p_{\rho} \p_{\rho'} G_R (\rho, \eta; \rho', \eta')$ with $G_R (\rho, \eta; \rho', \eta')$ the {\it retarded} propagator for a massless field in the Rindler spacetime  with Dirichlet boundary condition at $\rho =1$.
Fourier transforming (\ref{eomC}) to momentum space and using
$\int d\eta \, e^{i \nu \eta} \tilde G_R (\eta)
 = i \nu$,
 one finds
 that (\ref{SchE}) is modified to
 \be
 \hat H (\vec k, l=0) \psi - {i \om n_{\pm} \ov 4 \pi^3 \apr \mu_7 A}  \delta (\th) \psi (\th)= \om^2 \psi (\th)
 \ee
with $A = \sqrt{-g} (-g^{tt})$.
Writing the dispersion relation as $\om = \om_n - {i \ov 2} \Ga_n$ where $n$ denotes the excitation number, and using first order perturbation theory
in $n_{\pm}$ we find~\footnote{We normalize the eigenfunctions $\psi_n$ of $\hat H (\vec k, l)$ as
  $
  {1 \ov L^2 R^4} \int_0^{\pi \ov 2} d \th \, \sqrt{-g} \, G_{\phi \phi} \, (-g^{tt}) \, |\psi_n (\vec k, l; \th)|^2 = 1
  $ so that $\psi_n$ is dimensionless and has a smooth zero temperature limit.
  It depends on ratios of $T,M$ and $k$ but not explicitly on $\lam$.
  }
 \be \label{widF}
 \Ga_n^{(\pm1)} = {32 \pi^3 \sqrt{\lam} \ov N_c  m_q^2} |\psi_n (\th =0)|^2 \, n_\pm \
 \ee
with $\psi_n (\th=0)$ eigenfunctions of (\ref{SchE}) evaluated at the tip of the brane.
 Recall that $n_\pm$ are thermal densities for quarks and antiquarks and are proportional to $N_c$.

\begin{figure}[h!]
\vspace{-.2cm}
\centerline{\hbox{\psfig{figure=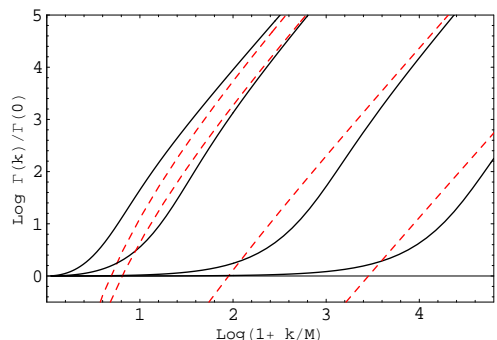,width=7cm,height=4.5cm}}}
\vspace{-.4cm}
\caption{The behavior of the width as a function
of $k$ for
$T/T_{\rm{diss}}=.99,\, .71,\, .3,\, .13$
from left to right. The dashed lines are analytic results for large momenta.
\label{fig:2}
}
\end{figure}

The ratio
 \be \label{eq:ratio}
 {\Ga_n (k) \ov \Ga_n (0)} = {|\psi_n (\th =0; \vec k)|^2 \ov |\psi_n (\th =0; \vec k=0)|^2} \
 \ee
can be evaluated numerically and the results are shown in
Fig.~\ref{fig:2}. For large $k$, the asymptotic form of the wave function, found in \cite{Ejaz:2007hg}, can be used to show that the width (\ref{eq:ratio}) scales like $k^2$ for large $k$: $\Ga_n (k) / \Ga_n (0) = R_n[T/M] (k/M)^2 + \mathcal{O}(k)$ for
some function $R_n[T/M]$. Furthermore for temperatures $T\ll M$ and $k \gg {M^3 \ov T^2}$ one finds the closed form expression
${\Ga_n (k) \ov \Ga_n (0)}
\approx \frac{ 2 (4 \pi)^4}{(n+2) (n+3/2)} \frac{ T^4 k^2}{M^6}$.
Fig.~\ref{fig:2} has the interesting feature that the
width is roughly constant for small $k$, but turns up quadratically
around $k/M =0.52 (T_{\rm{diss}}/T)^2$, which is roughly the momentum
at which the meson approaches its limiting velocity $v_c (T)$.  This is
consistent with the conclusions based on the velocity
dependence of the screened quark potential found
in \cite{Liu:2006nn}. Note that the width as defined here
is in the rest frame of the plasma, so the
$k^2$ behavior at large $k$ should be contrasted with the $1/k$ behavior
of a relativistic decay width which comes from the
usual time-dilation argument.

The plots here also share some similarities with
those in~\cite{Myers:2008cj} for momentum-dependence of meson widths
obtained for $\mu>m_q^{(T)}$ with $\lambda=\infty$
where the relevant D7 brane embedding
resembles a long spike reaching down to the horizon.

\vspace{-0.2cm}

\vspace{-0.2cm}

\begin{figure}[h!]
\centerline{\hbox{\psfig{figure=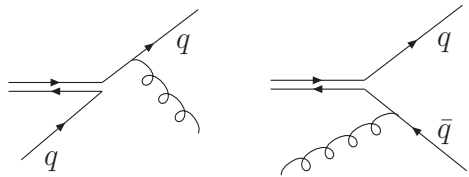,width=7cm,height=2.5cm}}}
\caption{\emph{Schematic} diagrams of the relevant
thermal processes contributing to the meson width.
For large $\lambda$ the first process is dominant,
coming from the single instanton sector.
\label{fig:3}}
\end{figure}

\vspace{-0.2cm}

Our result (\ref{widF}) has a very simple physical interpretation as shown in
the left plot of Fig.~\ref{fig:3}, in which a meson breaks apart by colliding with a quark (or anti-quark) in the thermal bath.
There are also processes corresponding to
a meson breaking up by colliding with gluons in the thermal bath,
shown in the right plot of Fig.~\ref{fig:3}. For such a process to happen the gluon should have an energy above the binding energy of the meson. The density of such gluons is thus suppressed by $e^{-2 \beta m_q^{(T)}}$ and should be described by
an instanton and anti-instanton so that the resulting worldsheet has trivial topology.
We expect that contributions from such processes are also controlled by
the value of the meson wave function at the tip of the brane,
and possibly
have similar growth with momentum.

Our method should be generic to any theory with a holographic dual.
While the precise value of the width is clearly model dependent,
it is conceivable that
the momentum dependence may apply in a wider context including QCD. In particular, our result highlights the contributions to quarkonium suppression from the collisions with medium quarks and gluons in the large transverse momentum region in heavy ion collisions.

Consider the effect of such a momentum scaling on $J/\psi$
with $M \approx 3 \rm{GeV}$.
The dissociation temperature from the gravity set-up is $T_{\rm{diss}} = 0.122M \approx 370 \rm{MeV}$ in fairly good
agreement with lattice data \cite{Asakawa:2003re}
for QCD $T_{diss} \approx 2.1 T_c$ for $T_c = 170
\rm{MeV}$~\cite{Mateos:2007vc}.
If we take the RHIC temperature of $T=250 \rm{MeV}$
(this corresponds to the second curve from the left in Fig.~2)
then a moving $J/\Psi$'s width will increase by
a factor of $2(10)$ at a momentum $k=6(13) \rm{GeV}$. When the width of a meson approaches the spacing between different meson states, the meson can no longer be considered as a well-defined quasi-particle. The momentum scaling thus implies a maximal momentum beyond which the meson no longer exists. As an illustration, suppose the width for the $J/\psi$ in the QGP at zero momentum is about 200 MeV (which is not known) then one expects a maximal momentum around 7 GeV.

Finally, we expect the worldsheet instantons found here to have many other applications to various aspects of flavor physics in AdS/CFT.

\begin{acknowledgments}

We thank C.~Athanasiou, H.~Meyer, K.~Rajagopal, D.~Teaney, A.~Tseytlin, U.~Wiedemann 
for useful discussions.
Research supported in part by
the DOE
under
contracts
\#DF-FC02-94ER40818.
HL is also supported
in part by the A.~P.~Sloan Foundation and the 
DOE OJI program.  

\end{acknowledgments}



\begin{thebibliography}{9}
%


\bibitem{Matsui:1986dk}
  T.~Matsui and H.~Satz,
  Phys.\ Lett.\ B {\bf 178}, 416 (1986).


\bibitem{Kharzeev:1994pz}
  D.~Kharzeev and H.~Satz,
  Phys.\ Lett.\  B {\bf 334}, 155 (1994)
  [arXiv:hep-ph/9405414].



\bibitem{AdS/CFT}
  J.~M.~Maldacena,
  Adv.\ Theor.\ Math.\ Phys.\  {\bf 2}, 231 (1998);
  E.~Witten,
 {\it ibid}. 505 (1998);
  S.~S.~Gubser, I.~R.~Klebanov and A.~M.~Polyakov,
  Phys.\ Lett.\ B {\bf 428}, 105 (1998).


\bibitem{Liu:2006nn}
  H.~Liu, K.~Rajagopal and U.~A.~Wiedemann,
  Phys.\ Rev.\ Lett.\  {\bf 98}, 182301 (2007)
  [arXiv:hep-ph/0607062];
  H.~Liu, K.~Rajagopal and U.~A.~Wiedemann,
  JHEP {\bf 0703}, 066 (2007)
  [arXiv:hep-ph/0612168].

\bibitem{Peeters:2006iu}
  K.~Peeters, J.~Sonnenschein and M.~Zamaklar,
  Phys.\ Rev.\  D {\bf 74}, 106008 (2006)
  [arXiv:hep-th/0606195];
  M.~Chernicoff, J.~A.~Garcia and A.~Guijosa,
  JHEP {\bf 0609}, 068 (2006)
  [arXiv:hep-th/0607089];




\bibitem{Karch:2002sh}
  A.~Karch and E.~Katz,
  JHEP {\bf 0206}, 043 (2002)
  [arXiv:hep-th/0205236].





\bibitem{Mateos:2007vn}
  D.~Mateos, R.~C.~Myers and R.~M.~Thomson,
  JHEP {\bf 0705}, 067 (2007)
  [arXiv:hep-th/0701132].


\bibitem{Ejaz:2007hg}
  Q.~J.~Ejaz, T.~Faulkner, H.~Liu, K.~Rajagopal and U.~A.~Wiedemann,
  JHEP {\bf 0804}, 089 (2008)
  [arXiv:0712.0590 [hep-th]].


\bibitem{Kruczenski:2003be}
  M.~Kruczenski, D.~Mateos, R.~C.~Myers and D.~J.~Winters,
  JHEP {\bf 0307}, 049 (2003)
  [arXiv:hep-th/0304032].
  J.~Babington, J.~Erdmenger, N.~J.~Evans, Z.~Guralnik and I.~Kirsch,
  Phys.\ Rev.\ D {\bf 69}, 066007 (2004)
  [arXiv:hep-th/0306018].
  M.~Kruczenski, D.~Mateos, R.~C.~Myers and D.~J.~Winters,
  JHEP {\bf 0405}, 041 (2004)
  [arXiv:hep-th/0311270].


\bibitem{Hoyos:2006gb}
  C.~Hoyos-Badajoz, K.~Landsteiner and S.~Montero,
  JHEP {\bf 0704}, 031 (2007)
  [arXiv:hep-th/0612169].


\bibitem{Mateos:2007vc}
  D.~Mateos, S.~Matsuura, R.~C.~Myers and R.~M.~Thomson,
  JHEP {\bf 0711}, 085 (2007)
  [arXiv:0709.1225 [hep-th]].

  S.~Kobayashi, D.~Mateos, S.~Matsuura, R.~C.~Myers and R.~M.~Thomson,
  JHEP {\bf 0702}, 016 (2007)
  [arXiv:hep-th/0611099].


\bibitem{Karch:2007br}
  A.~Karch and A.~O'Bannon,
  JHEP {\bf 0711}, 074 (2007)
  [arXiv:0709.0570 [hep-th]].

  K.~Ghoroku, M.~Ishihara and A.~Nakamura,
  Phys.\ Rev.\  D {\bf 76}, 124006 (2007)
  [arXiv:0708.3706 [hep-th]].

  S.~Nakamura, Y.~Seo, S.~J.~Sin and K.~P.~Yogendran,
  arXiv:hep-th/0611021.


\bibitem{Myers:2008cj}
  R.~C.~Myers and A.~Sinha,
  arXiv:0804.2168 [hep-th].

\bibitem{Erdmenger:2007ja}
  J.~Erdmenger, M.~Kaminski and F.~Rust,
  Phys.\ Rev.\  D {\bf 77}, 046005 (2008)
  [arXiv:0710.0334 [hep-th]].

  J.~Mas, J.~P.~Shock, J.~Tarrio and D.~Zoakos,
  arXiv:0805.2601 [hep-th].


\bibitem{Fradkin:1985qd}
  E.~S.~Fradkin and A.~A.~Tseytlin,
  Phys.\ Lett.\  B {\bf 163}, 123 (1985).

\bibitem{Dine:1986zy}
 M.~Dine, N.~Seiberg, X.~G.~Wen and E.~Witten,
 Nucl.\ Phys.\  B {\bf 278}, 769 (1986);
  E.~Witten,
  JHEP {\bf 0002}, 030 (2000)
  [arXiv:hep-th/9907041].


\bibitem{Asakawa:2003re}
  M.~Asakawa and T.~Hatsuda,
  Phys.\ Rev.\ Lett.\  {\bf 92}, 012001 (2004);
  S.~Datta,
 F.~Karsch, P.~Petreczky and I.~Wetzorke,
  Phys.\ Rev.\ D {\bf 69}, 094507 (2004).
 R.~Morrin {\it et al.},
  PoS {\bf LAT2005}, 176 (2005).
  G.~Aarts, C.~Allton, M.~B.~Oktay, M.~Peardon and J.~I.~Skullerud,
  Phys.\ Rev.\  D {\bf 76}, 094513 (2007)
  [arXiv:0705.2198 [hep-lat]].



\end{thebibliography}
\end{document}